\documentclass[aps, pra, showkeys, showpacs, reprint]{revtex4-1} 

\usepackage[usenames,dvipsnames]{color} 
\usepackage{amsmath} 
\usepackage{txfonts} 
\usepackage{upgreek} 
\usepackage{todonotes} 
\usepackage{amssymb} 
\usepackage{braket} 
\usepackage{array} 

\usepackage[unicode]{hyperref}
\hypersetup{
    unicode=true,                                           
    a4paper=true,
    plainpages=false,
    pdffitwindow=true,                                      
    pdftitle={\textsc{CooperativeLatticePaper}},                                
    pdfauthor={Robert J. Bettles},            
    pdfsubject={},                                             
    pdfkeywords={dipole-dipole, subradiance, cooperative},
    colorlinks=true,                                        
    linkcolor=blue,                                         
    citecolor=blue,                                         
    filecolor=blue,                                         
    urlcolor=blue                                           
}
\usepackage{upgreek} 

\usepackage[caption=false]{subfig} 
\usepackage{graphicx} 
\usepackage[countmax]{subfloat}
\usepackage{import} 
\usepackage{transparent} 

\definecolor{MyGreen}{rgb}{0,0.5,0} 

\newcommand{\rvec}{\ensuremath{\mathrm{\mathbf{r}}}} 
\newcommand{\bvec}[2]{\ensuremath{\boldsymbol{\mathrm{#1}}_{#2}}} 

\usepackage{tikz}

\usepackage{soul}  

\begin{document}

\bibliographystyle{apsrev4-1}

\title{Cooperative ordering in lattices of interacting two-level dipoles}
\author{Robert J. Bettles}\email{r.j.bettles@durham.ac.uk}
\author{Simon A. Gardiner}\email{s.a.gardiner@durham.ac.uk}
\author{Charles S. Adams}\email{c.s.adams@durham.ac.uk}
\affiliation{Joint Quantum Center (JQC) Durham--Newcastle, Department of Physics,  Durham University, South Road, Durham, DH1 3LE, United Kingdom}
\date{\today}

\begin{abstract}
We investigate the cooperative behavior of regular monolayers of driven two-level dipoles, using classical electrodynamics simulations. The dipolar response results from the interference of many cooperative eigenmodes, each frequency-shifted from the single resonant dipole case, and with a modified lifetime, due to the interactions between dipoles. Of particular interest is the kagome lattice, where the semiregular geometry permits simultaneous excitation of two dominant modes, one strongly subradiant, leading to an electromagnetically-induced-transparency-like interference in a two-level system. The interfering modes are associated with ferroelectric and antiferroelectric ordering in alternate lattice rows with long range interactions.  
\end{abstract}


\maketitle

\section{Introduction}

\begin{figure*}[t]
\centering
{\includegraphics[width=1\textwidth]{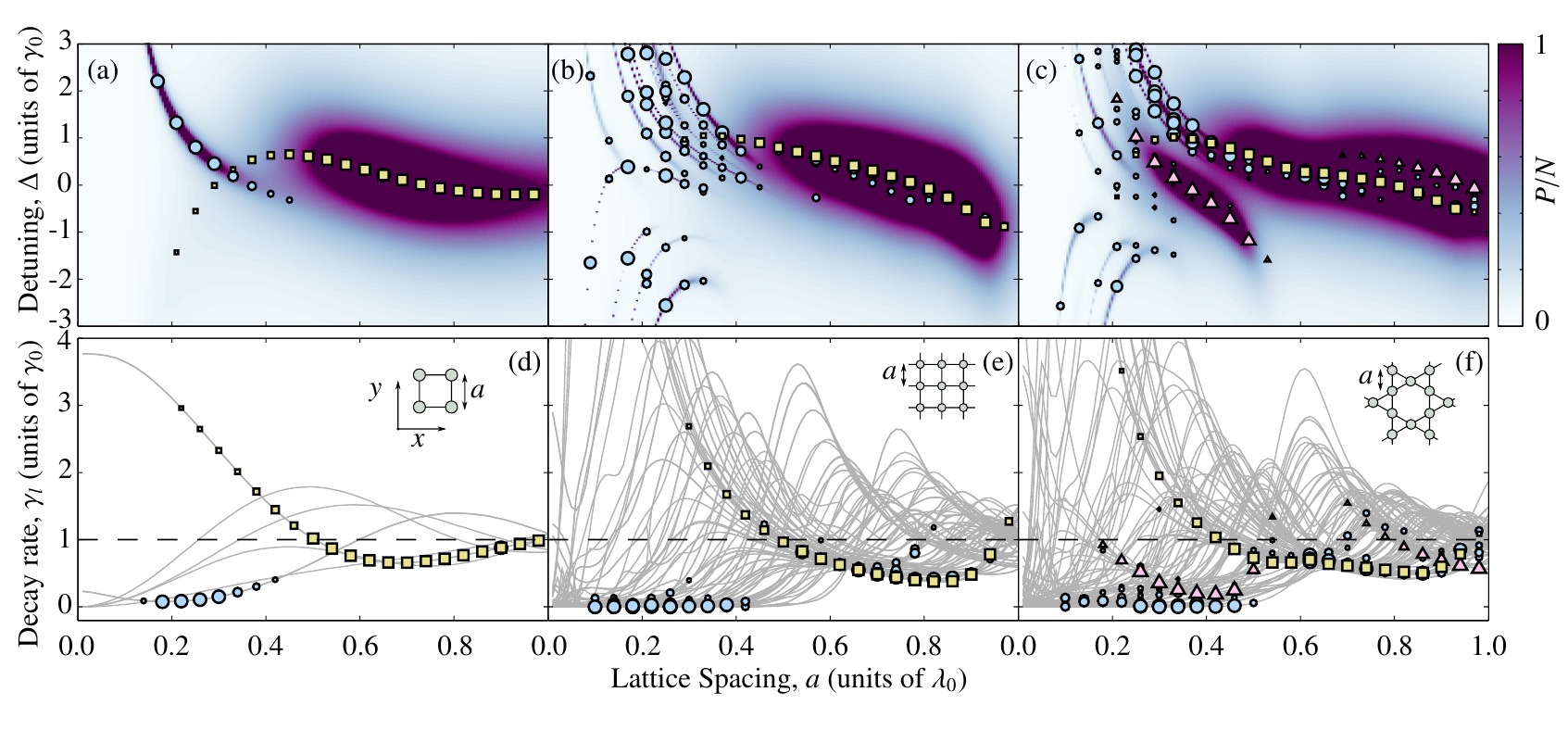}}
\caption{(Color online) Scattered power and mode eigenvalues for: [(a) and (d)] a square lattice with $N=2\times 2=4$ driven interacting dipoles, [(b) and (e)] a square lattice with $N=7\times 7=49$, and [(c) and (f)] a kagome lattice with $N=47$. Snapshots of the lattice structure are shown inset in (d)--(f). (a)--(c) Color scale shows the total scattered power relative to the scattered power from $N$ non-interacting resonant dipoles, $P/N=\sum_i \bvec{d}{i}^*\cdot\bvec{d}{i}/N\alpha_0^2 E_0^2$ [unitless, see Eq.~(\ref{eq:ScattPower})]. Dipoles are positioned in the $xy$ plane with nearest-neighbor spacing $a$ and irradiated by a $y$-linearly-polarized uniform light beam, propagating in $z$ and detuned from the dipole transition frequency by $\Delta$. In addition, we plot the real eigenvalue components (shifts) for all modes with $|c_{l}|^{2}>0.1$ (marker size $\propto |c_{l}|^{2}$). (d)--(f) Imaginary eigenvalue components (decay rates, $\gamma_l$) for all eigenmodes (grey lines); as in (a)--(c), those eigenmodes with $|c_{l}|^{2}>0.1$ are highlighted with scatter points.} 
\label{Fig:1}
\end{figure*}

Coherent emission from an ensemble of scatterers (such as electric dipoles) can result in the scatterers behaving as a collective rather than independently \cite{Dicke1954}. Such ``cooperative'' emission can lead to large frequency shifts off-resonance, and to dramatically modified decay rates (superradiance and subradiance) \cite{Stephen1964,Lehmberg1970,Friedberg1973}. This has been realized experimentally in a number of systems, including ions \cite{DeVoe1996,Meir2014,Casabone2015}, nuclei \cite{Rohlsberger2010}, quantum dots \cite{Scheibner2007}, nanoplasmonics \cite{Luk'yanchuk2010}, Bose-Einstein condensates \cite{Inouye1999} and both room temperature \cite{Keaveney2012} and cold atoms \cite{Yoshikawa2005,Greenberg2012,Goban2015a}. Other related cooperative phenomena include highly directional scattering \cite{Rouabah2014}, excitation localization \cite{Jenkins2012,Akkermans2008,Skipetrov2015}, and modified optical transmission and scattering \cite{Chomaz2012,Pellegrino2014a,Kemp}.

Cooperative emission is caused by the interference of radiation from individual scatterers, and periodic spacing between neighboring dipoles can therefore lead to a significantly enhanced cooperative response \cite{Meir2014,Jenkins2012,Nienhuis1987,Chang2012,Olmos2013}. Coherent scattering between two-level dipoles maps exactly onto a spin exchange description \cite{Yan2013,Barredo2015}; consequently, there is a unifying crossover between cooperative light scattering and interacting spin systems. Spin lattices are a subject of widespread contemporary interest, and manifest in such diverse systems as quantum degenerate gases \cite{Bakr2009,Weitenberg2011}, polar molecules \cite{Micheli2006,Yan2013} and cold atoms \cite{Nogrette2014,Gonzalez-Tudela2015} in optical lattices, and electric and magnetic multipoles in plasmonic nanostructures \cite{Liu2009a,Luk'yanchuk2010,Fedotov2010,Rougemaille2011}. An understanding of the cooperative behavior in these driven-dissipative systems could open the door to a wide range of applications (e.g.\ shifts and lifetimes in optical lattice clocks \cite{Kramer}, narrow linewidth superradiant lasers \cite{Bohnet2012,Ludlow2015}, subwavelength light control \cite{Jenkins2012} and many body spin models \cite{Olmos2013,Yan2013}). 

One particular 2D lattice geometry associated with a range of exotic spin phenomena is the trihexagonal (kagome) lattice. Examples include spin ice and geometric frustration \cite{Moessner2006,Maksymenko2015}, photonic flat bands and band gaps \cite{Vicencio2014}, low-loss transmission through hollow-core photonic crystal fibers \cite{Couny2007}, and non-integer Mott phases in optical lattices \cite{Parameswaran2013,Zhang2015a}. In this work we show that the semiregular geometry of kagome lattices makes it possible to straightforwardly populate cooperative states associated with dramatic interference line-shapes and strongly subradiant modes. 

\section{Interacting Dipole Model}
We calculate the cooperative shifts and decay rates in periodic 2D monolayers of interacting dipoles, using a model closely following that of \cite{Morice1995,Ruostekoski1997a,Javanainen1999,Chomaz2012,Jenkins2012}. We treat each dipole as a weakly-driven damped oscillator, with electric dipole moment $\bvec{d}{i}=\alpha \bvec{E}{}(\bvec{r}{i})$ ($i\in 1,\dots,N$) proportional to the total incident electric field $\bvec{E}{}(\bvec{r}{i})$ and the polarizability $\alpha$. The dipole positions $\bvec{r}{i}$ form a 2D lattice in the $xy$ plane with nearest-neighbor spacing $a$. 
In this work we treat the dipoles as 2-level atoms. Such atomic lattices could be realized in a Mott-insulator phase in an optical lattice \cite{Greiner2002,Weitenberg2011} or dipole trap array \cite{Nogrette2014,Lester2015a}. Dipolar 2D lattices of polar molecules \cite{Yan2013} and plasmonic nanoresonators \cite{Liu2009a,Luk'yanchuk2010,Fedotov2010,Rougemaille2011} have also been demonstrated, and in these systems we would expect similar qualitative behavior to the results in this work. The effects of finite potential trap depths \cite{Jenkins2012} and imperfect filling (vacant lattice sites) will be addressed in future work.

For a 2-level $J=0\to J=1$ atomic transition (e.g.\ Sr \cite{Olmos2013,Stellmer2014}), the polarizability takes the form ${\alpha = -\alpha_{0}\gamma_{0}/(\Delta+\mathrm{i}\gamma_{0})}$, where $\Delta = \omega-\omega_0$ is the detuning of the electric field frequency $\omega$ from resonance, $\gamma_{0}$ is the vacuum coupling or scattering rate, and $\alpha_{0} = 6\pi\varepsilon_{0}/k_{0}^{3}$ (SI units) quantifies the magnitude of the polarizability on-resonance (the wavenumber $k_0$ corresponds to the resonant wavelength $\lambda_0 = 2\pi/k_0$, and $\varepsilon_0$ is the vacuum permittivity). We take the driving field $\bvec{E}{0}$ to be a plane wave of amplitude $E_{0}$ propagating along $z$ and linearly polarized in $y$.  Each dipole also radiates a secondary electric field, hence the total field felt by the $i$th dipole, $\bvec{E}{}(\bvec{r}{i}) = \bvec{E}{0}(\bvec{r}{i}) + \sum_{j\neq i} \bvec{E}{j}(\bvec{r}{i})$, is the sum of the applied driving field $\bvec{E}{0}$ and the fields $\bvec{E}{j}$ radiated from all other dipoles. The field radiated by the $j$th dipole is proportional to its dipole moment $\bvec{E}{j}(\rvec) = \mathop{\mathsf{G}(\rvec-\bvec{r}{j})} \bvec{d}{j}$, where $\mathsf{G}$ is the dipole propagation tensor (as given in \cite{Jackson1963}). The matrix elements in a Cartesian representation of  $\mathsf{G}$ are given by
\begin{equation}\label{eq:Gtensor}
\mathsf{G}_{p,q}(\rvec) = \frac{1}{\varepsilon_0} \left[ \left( \frac{\partial}{\partial r_{p}} \frac{\partial}{\partial r_{q}}   - \delta_{p,q} \nabla^2 \right) \frac{\mathrm{e}^{\mathrm{i}k_0 r}}{4\pi r} - \delta_{p,q}\delta(\rvec)  \right],
\end{equation}
where $p,q\in \{1,2,3\}$, $\{r_{1}, r_{2}, r_{3}\}$ are the components of $\mathbf{r}$ directed along the $\{\hat{\boldsymbol{x}}, \hat{\boldsymbol{y}}, \hat{\boldsymbol{z}}\}$ unit vectors, $r=|\mathbf{r}|$, $\delta_{p,q}$ is a Kronecker delta, and $\delta(\mathbf{r})$ is a Dirac delta function.

Substituting $\bvec{E}{j}(\rvec)$ into the expression for $\bvec{d}{i}$ yields a system of coupled linear equations,
\begin{equation}\label{eq:dtot}
\bvec{d}{i} = \alpha \bvec{E}{0}(\bvec{r}{i}) + \alpha \sum_{j\neq i} \mathop{\mathsf{G}(\bvec{r}{i}-\bvec{r}{j})} 
\bvec{d}{j}.
\end{equation}
Eq.~(\ref{eq:dtot}) is shown in \cite{Svidzinsky2010} to be equivalent to treating the dipoles both quantum mechanically (assuming weak excitation) or as classical harmonic oscillators.
Following the method in \cite{Chomaz2012}, we solve Eq.~(\ref{eq:dtot}) by writing it as a matrix equation, 
$\vec{\bvec{E}{}}_0 = \mathop{\boldsymbol{\mathsf{M}}} \vec{\bvec{d}{}}$, determining the inverse matrix $\boldsymbol{\mathsf{M}}^{-1}$ numerically and then solving for $\vec{\bvec{d}{}}$. Here $\vec{\bvec{E}{}}_{0}$ and $\vec{\bvec{d}{}}$ are dimensionless column vectors of $\bvec{E}{0}(\lambda_{0}\tilde{\boldsymbol{\mathrm{r}}}_{i})/E_{0}$ and $\bvec{d}{i}/\alpha_{0}E_{0}$, respectively ($\tilde{\boldsymbol{\mathrm{r}}}_{i}=\boldsymbol{\mathrm{r}}_{i}/\lambda_{0}$ is a dimensionless position vector), and $\boldsymbol{\mathsf{M}}$ is a dimensionless $3N\times 3N$ matrix describing all the driving and coupling terms, with elements of the form
$\alpha_{0}\{\alpha^{-1}\delta_{p,q}\delta_{i,j} 
- \mathsf{G}_{p,q}(\lambda_{0}[\tilde{\boldsymbol{\mathrm{r}}}_{i}-\tilde{\boldsymbol{\mathrm{r}}}_{j}])\}$. 

\section{Cross-Section and Scattered Power}
It is instructive to decompose the vectors $\vec{\bvec{E}{}}_0 = \sum_l \mu_{l}c_{l} \vec{\bvec{m}{}}_{l}$ and $\vec{\bvec{d}{}} = \sum_l c_{l}\vec{\bvec{m}{}}_{l}$ in terms of the eigenvectors, $\vec{\bvec{m}{}}_{l}$, of $\boldsymbol{\mathsf{M}}$  (with corresponding eigenvalues $\mu_l$) \cite{Hopkins2013}.
The coefficients $c_{l}$ can be calculated by projecting $\vec{\bvec{d}{}}$ onto $\vec{\bvec{m}{}}_{l}$. Note that, while  $\mu_{l}$ and $c_{l}$ both depend on the polarizability $\alpha$ (and hence the detuning $\Delta$), the products $\mu_{l}c_{l}$ are independent of $\alpha$, as are the eigenvectors $\vec{\bvec{m}{}}_{l}$ 
\footnote{Eigenvectors for degenerate eigenvalues (e.g.\ in the square lattice with $x-y$ symmetry) are not uniquely defined since a linear combination of these eigenvectors also has the same eigenvalue. This is of minor importance for this work.}.
The matrix $\boldsymbol{\mathsf{M}}$ is not Hermitian but rather complex symmetric; such matrices commonly appear in scattering problems, e.g.\ in nanoparticle plasmonics \cite{Hopkins2013}, multi-photon ionization \cite{Jaron2000}, and cold atoms \cite{Svidzinsky2008,Svidzinsky2010,Bellando2014}. The non-Hermiticity results in non-orthogonal eigenvectors, and hence in interference terms appearing in the (dimensionless) total scattered power $P$ and extinction cross-section $\sigma$: 
\begin{align} 
\label{eq:ScattPower}
P
=
\vec{\bvec{d}{}}^{*}\cdot\vec{\bvec{d}{}} = &
\sum_{l}
\left( |c_{l}|^2 + \sum_{k\neq l} c_{l}^{*} c_{k} \vec{\bvec{m}{}}_{l}^{*} \cdot \vec{\bvec{m}{}}_{k}    \right),
\\
\label{eq:ExtCrossSection}
\sigma 
=
\mathop{\textrm{Im}} \left( \vec{\bvec{E}{}}_0^* \cdot \vec{\bvec{d}{}} \right) 
 = & \mathop{\textrm{Im}} \left( \sum_{l}  \mu_{l}^{*}
 \left[ |c_{l}|^2 + \sum_{k\neq l} c_{l}^{*} c_{k} \vec{\bvec{m}{}}_{l}^{*} \cdot \vec{\bvec{m}{}}_{k}    \right]
 \right).
\end{align}
Each of the direct sum terms in Eq.~(\ref{eq:ExtCrossSection}) can be approximated by a Lorentzian line shape:
\begin{equation} \label{eq:LorentzianLineshape}
|c_{l}|^2 \mathop{\textrm{Im}} (\mu_{l}^{*}) =-|c_{l}|^2 \mathop{\textrm{Im}} (\mu_{l}) = f_l \frac{\gamma_l^2}{(\Delta-\Delta_l)^2 + \gamma_l^2},
\end{equation}
where $f_{l}$ is the peak of the line-shape (attained when $\Delta=\Delta_{l}$), $\Delta_{l}/\gamma_{0} = \mathop{\textrm{Re}}(\mu_{l})$ is the line-centre and $\gamma_{l}/\gamma_{0}=-\mathop{\textrm{Im}}(\mu_{l})$ is the half-width, relative to the vacuum coupling. Relating the half-width to the characteristic decay lifetime $\tau \sim (2\gamma_0)^{-1}$, the imaginary parts of the eigenvalues $\mu_{l}$ can lead to superradiance ($\gamma_l/\gamma_0>1$) and subradiance ($\gamma_l/\gamma_0<1$).

\begin{figure}[t]
\centering
{\includegraphics[width=0.48\textwidth]{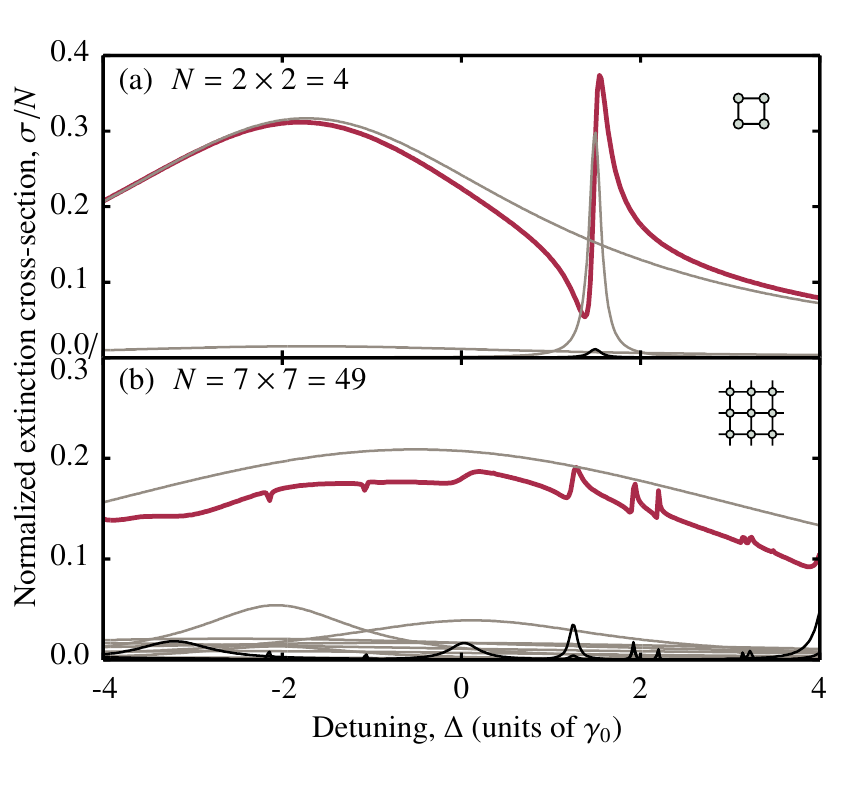}}
\caption{(Color online) Fano resonances and interferences in the extinction cross-section, through square dipolar lattices with (a) $N=4$ and (b) $N=49$, and with lattice spacing $a=0.2\lambda_0$. The thick red lines show the extinction cross-section $\sigma/N$ [unitless, see Eq.~(\ref{eq:ExtCrossSection})] as a function of the detuning of the driving light relative to the scattering rate, $\Delta$. The thin grey lines are the calculated direct terms $\mathop{\textrm{Im}}(\mu_l^*)|c_l|^2/N$ in Eq.~(\ref{eq:ExtCrossSection}) associated with each eigenmode. These can be well approximated as Lorentzian line-shapes [Eq.~(\ref{eq:LorentzianLineshape})]. Narrower weak modes are highlighted in black for clarity, and are typically associated with small Fano-resonance-like features in the extinction cross-section $\sigma$.} 
\label{Fig:2}
\end{figure}

In Fig.~\ref{Fig:1}(a--c) we plot the normalized power ${P/N \equiv \sum_{i} {\bvec{d}{i}^{*}\cdot\bvec{d}{i}}/N\alpha_{0}^{2}E_{0}^{2}}$ (where $N$ is the total number of dipoles) scattered from square and kagome lattices. We characterize the relative contribution of each mode by $|c_{l}|^{2}$ 
\footnote{Once $|\Delta _{l}|>3\gamma _l$ for a particular mode, $f_{l}$ and $|c_{l}|^{2}$ are no longer calculated since this is outside the calculated parameter range of $\protect \mathaccentV {vec}17E{\protect \ensuremath {\protect \boldsymbol {\protect \mathrm {d}}_{}}}$.}.
Highlighting modes with $|c_{l}|^{2}>0.1$ (scatter points), we see how the overall scattering behavior of the lattice (color scale) is due to the simultaneous population of several different eigenmodes, each with its own behavior as determined by its eigenvalues.  In Fig.~\ref{Fig:1}(d--f) we plot the corresponding widths $\gamma_l$ for all eigenmodes of $\boldsymbol{\mathsf{M}}$ (gray lines). Looking at the general behavior of these eigenmodes as well as the selection of modes populated by our choice of driving field, we observe certain similarities between these lattices (as well as with similarly-sized triangular and hexagonal lattices not shown here). In all three lattices, we highlight a similar-looking mode (yellow squares) which at small $a$ is superradiant and red-shifted. This is the spin analogue of the Dicke symmetric state \cite{Dicke1954} with $\gamma_l/\gamma_0 \simeq N$, which we will later show corresponds to having all spins aligned with the field, i.e.\ a ferroelectric-like state. Similarly, each lattice exhibits strongly subradiant modes (blue circles) which, like the ``yellow squares'' mode, are shifted off-resonance as $a\to 0$, due to the $1/r^3$ small $r$ behavior of $\mathop{\textrm{Re}}(\mathsf{G})$ in Eq.~(\ref{eq:Gtensor}) \cite{Jackson1963}. Comparing Fig.~\ref{Fig:1}(a) and \ref{Fig:1}(b) we see that the overall behavior of the square lattices is broadly similar, barring the introduction of more modes in \ref{Fig:1}(b). In contrast, with a kagome lattice [Fig.~\ref{Fig:1}(c)] a pronounced new structure appears (pink triangles). We devote the rest of this paper to explaining this structure and how it combines aspects of cooperative electromagnetically-induced transparency (EIT) in two-level systems with combined ferroelectric and anti-ferroelectric responses in spin systems.

\section{Square and Kagome Lattice Cross-Sections}
In Fig.~\ref{Fig:2} we plot the normalized extinction cross-section $\sigma/N \equiv \sum_{i} \mathop{\textrm{Im}} [\bvec{E}{0}^{*}(\bvec{r}{i}) \cdot \bvec{d}{i}]/N\alpha_{0}E_{0}^{2}$ through the same $N=4$ and $N=49$ square lattices as in Fig.~\ref{Fig:1}(a,b), with lattice spacing $a=0.2\lambda_0$. For the small $N=4$ lattice (a), we observe one strong broad red-shifted mode and one strong narrow blue-shifted mode (as well as a few much weaker modes). Where the two modes overlap there is a strong asymmetric resonance in the cross-section line-shape. This Fano-like resonance is due to the interference terms that appear in Eq.\ (\ref{eq:ExtCrossSection}) and is a direct consequence of the non-zero overlap between mode vectors (the eigenvector non-orthogonality). In the power spectrum for $N=49$ [Fig.~\ref{Fig:1}(b)], we observed that adding more dipoles to the square lattice resulted in many strong narrow modes appearing at small lattice spacings. These modes are also visible in the extinction cross-section [Fig.~\ref{Fig:2}(b)], however their relative contribution to the total line-shape is much weaker. As in the $N=4$ case, the line-shape for $\sigma/N$ when $N=49$ is dominated by one broad mode, but the asymmetric Fano resonances resulting from overlap with the narrower modes are much smaller. In a real experiment, with associated lattice imperfections or noise, these weak narrow modes will wash out (see Section \ref{sec:LatticeImperfections}).

\begin{figure}[t]
\centering
{\includegraphics[width=0.48\textwidth]{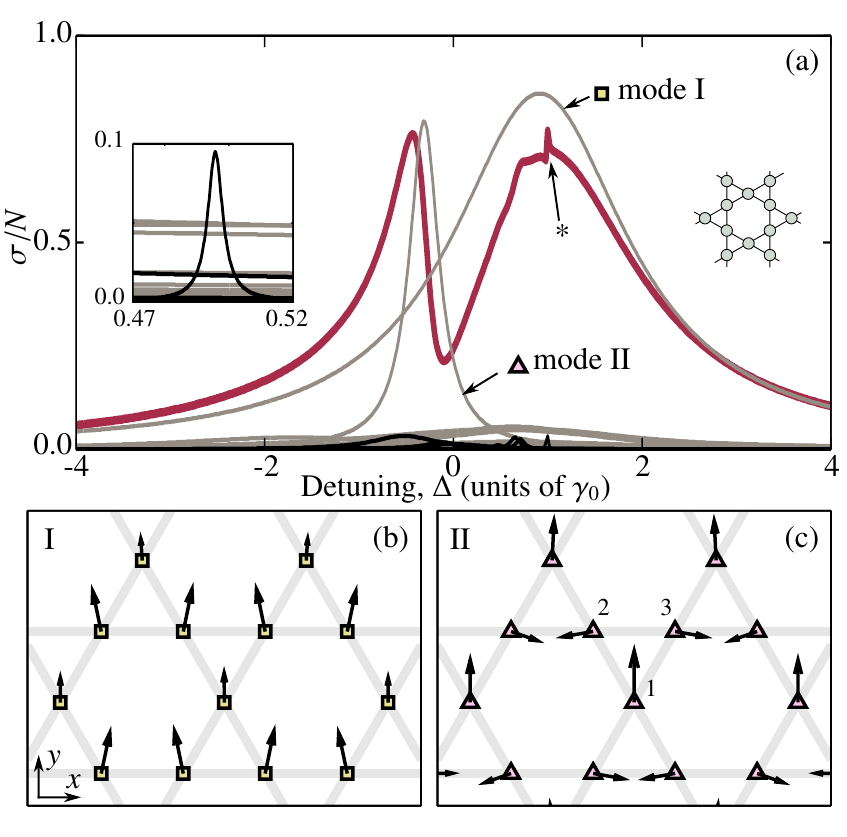}}
\caption{(Color online) Two-level cooperative EIT in the kagome lattice. (a) As in Fig.~\ref{Fig:2}, we plot the extinction cross-section $\sigma/N$ (Eq.~\ref{eq:ExtCrossSection}, unitless) as a function of detuning, now through a kagome lattice with $N=47$ and lattice spacing $a=0.4\lambda_0$ (thick red line). The thin grey lines are the calculated direct terms $\mathop{\textrm{Im}}(\mu_l^*)|c_l|^2/N$ in Eq.~(\ref{eq:ExtCrossSection}) associated with each eigenmode. Some of these are highlighted black to stand out. We label two particular modes (I and II), which correspond to the modes plotted in (b) and (c) respectively. The inset highlights the weak mode causing the interference labeled (*).
(b) and (c) show the $(x,y)$ vector components of the eigenmodes highlighted in yellow and pink respectively in Fig.~\ref{Fig:1}(c,f) and also correspond to the modes labeled I and II in (a) respecively. We plot the real, in-phase components of the eigenmodes at each lattice point on the kagome lattice. (b) highlights the ferroelectric mode and (c) the mixed-behavior strong subradiant mode. The lattice sites labeled 1, 2 and 3 are discussed in the text.} 
\label{Fig:3}
\end{figure}

For a kagome lattice geometry 
\footnote{We also performed calculations for triangular and hexagonal lattices, although only the kagome exhibited strong mode interferences like those shown here.}
however, we observe two very strong modes at lattice spacing $a=0.4\lambda_0$. As in the $2\times 2$ structure in Fig.~\ref{Fig:2}(a), these result in a distinctive interference line-shape in the extinction cross-section $\sigma$ [Fig.~\ref{Fig:3}(a)]. Similar to the $7\times 7$ square lattice case, most of the narrow modes prominent in the power spectrum [Fig.~\ref{Fig:1}(c)] are relatively weak in the extinction. The separate mode highlighted in Fig.~\ref{Fig:1}(c,f) with triangular pink markers remains significant, however, and interference between this mode (labeled II) and the broader strong mode (labeled I) results in a line-shape similar in appearance to those of cooperative and dipole EIT \cite{Waks2006,Zhang2008,Liu2009a,Jenkins2013,Puthumpally-Joseph2014a}. In these systems, interferences between different excitation modes (typically one narrow and one broad) result in transparency where otherwise extinction would be expected. This is analogous to conventional EIT, but here the excited states are the cooperative states of an ensemble of 2-level scatterers, as opposed to multiple states in a single scatterer. Compared to the square lattice in Fig.~\ref{Fig:2}(b), the strength of the two modes producing this transparency feature means it should be significantly more robust to experimental limitations and noise.

\section{Eigenmode Behavior}
To better understand the origin of these modes and why they are populated in the kagome lattice and not the square, triangular or hexagonal lattices, we analyze the eigenvectors themselves. In Fig.~\ref{Fig:3}(b,c) we plot the real $x$ and $y$ vector components of two important kagome eigenmodes (the imaginary components are much weaker). As in \ref{Fig:3}(a) we consider lattice spacing $a=0.4\lambda_0$, however the change in these eigenvectors over the parameter range of interest ($0.3<a/\lambda_0<0.5$) is negligible. The behavior of each mode is dependent only on the matrix $\boldsymbol{\mathsf{M}}$ and contains no information about the driving field polarization or geometry (except for the detuning $\Delta$ which appears in $\alpha$). The choice of driving field simply determines which modes are populated 
\footnote{Despite being limited computationally to lattices with $N\sim 50$, the mode vector patterns for lattices with larger atom numbers are still similar to those in Fig.~\ref{Fig:3}(b,c)}.
%
The strong broader mode [I: the yellow squares mode in Fig.~\ref{Fig:1}(b)] behaves ferroelectrically, with all vectors tending to align with the driving field along $y$ (c.f.\ the ferrimagnetic modes in \cite{Maksymenko2015}). A similar mode exists with all vectors aligned along $x$, but this doesn't couple with the chosen driving field polarization. As already mentioned, this is analogous to the symmetric Dicke state and appears in all the other lattices we have mentioned as well (square, triangular, hexagonal). 

Fig.~\ref{Fig:3}(c) shows the mode responsible for the strong interference line-shape [II: the pink triangles mode in Fig.~\ref{Fig:1}(c)]. In this mode we observe alternating rows of ferroelectric dipoles aligned with the driving field along $y$ (lattice site 1), and antiferroelectric dipoles perpendicular to the driving field and anti-aligned with their nearest neighbors (sites 2 and 3).
The long range nature of the dipole-dipole interaction combined with the non-trivial kagome geometry makes unraveling the origin of this mode behavior a complicated task. 
We can however gain insight through considering the individual contributions of different dipoles.
Considering first the dipole at lattice site 1, the dipole vectors of the nearest neighbors at sites 2 and 3 are symmetric in $x$, meaning the sum of the electric fields they radiate onto site 1 has only a $y$ component (the $x$ components cancel). The same is true for the remaining dipoles along row 2-3 and other rows of that type: for every dipole there is an equal and opposite mirror dipole along the same row canceling all the $x$ field components felt at site 1. The dipoles along the same row as dipole 1 contribute fields along $y$ as does the driving field, resulting in an overall dipole orientation along $y$ for dipole 1. 
Similar symmetries can be used to explain the behavior of the dipoles at sites 2 and 3, however what is striking is the stripe-like behavior of these alternating rows. 
The kagome lattice can be constructed by removing a triangular lattice with lattice period $2a$ from a triangular lattice with period $a$ \cite{Jo2012} and this double periodicity is manifest in mode II ($a$ spacing between antiferroelectric dipoles; $2a$ spacing between ferroelectric dipoles).
This suggests the mode is related to this double periodicity, which doesn't exist in the regular lattices. Furthermore, the kagome lattice can be classed as `semiregular', in that its tiling consists of triangular and hexagonal tiles surrounding common vertices, and so even though it shares the same common base unit tiles as the triangular and hexagonal lattices individually, its behavior is still significantly different.
It will be interesting to model the cooperative behavior of other semiregular geometries searching for similar features, as well as investigating the links between our kagome spin lattices and other semiregular lattice phenomena such as photonic flat bands \cite{Vicencio2014,Vicencio2015,Mukherjee2015} and geometric frustration \cite{Moessner2006,Maksymenko2015}.

Note that we have been describing the bulk mode behavior. The dipole orientations differ at the lattice edges since the contributions from nearby neighbors are different. However it is the bulk behavior that is characteristic of the modes in this paper and moving to larger lattices simply extends the region over which the bulk behavior manifests without significant changes to the behavior itself.

\begin{figure}[t]
\centering
{\includegraphics[width=0.48\textwidth]{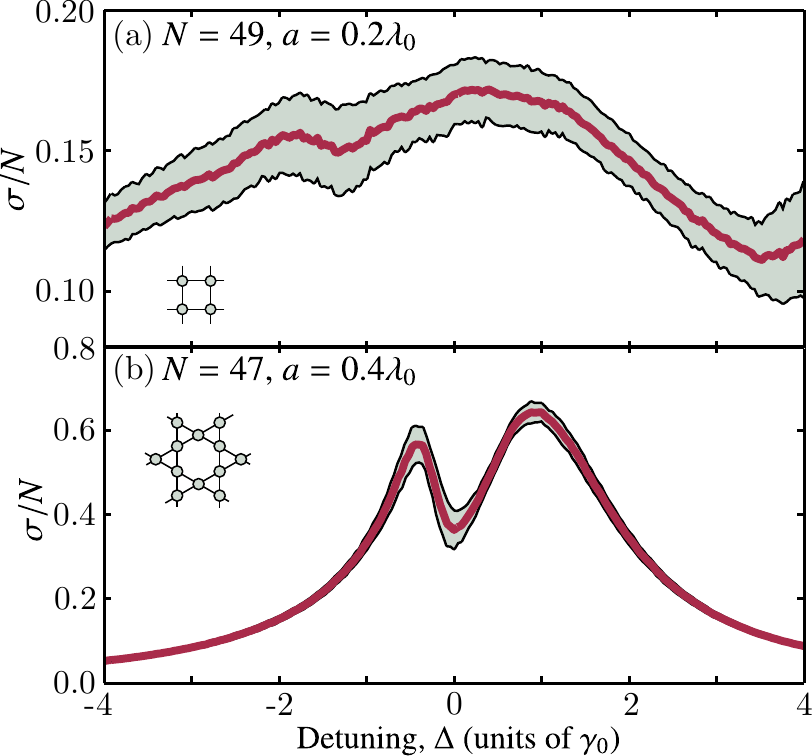}}
\caption{(Color online) The effect of experimental imperfections on extinction cross-section, $\sigma/N$ (Eq.\ \ref{eq:ExtCrossSection}, unitless), through (a) a square lattice with $N=49$ sites and lattice spacing $a=0.2\lambda_0$ and (b) a kagome lattice with $N=47$ sites and lattice spacing $a=0.4\lambda_0$. These correspond to the same lattice parameters as in Fig.\ \ref{Fig:2}(b) and Fig.\ \ref{Fig:3}(a) respectively. The red solid lines show the average over several hundred realizations; the black solid lines bounding the shaded areas represent the standard deviation. In each realization (and at each detuning), we remove at random 5 atoms from both lattices and sample the individual atom positions using a Gaussian distribution, modeling the effect of finite trap depth ($V_0=750 E_R$, where $E_R$ is the lattice recoil energy). 
} 
\label{Fig:4}
\end{figure}

\section{Lattice Imperfections} \label{sec:LatticeImperfections}
Finally, we consider the effect of experimental imperfections on the observed lineshapes. So far we have considered perfect systems where every lattice site is occupied by one atom centered exactly on that lattice site (assuming an infinite trapping potential). Here we calculate how some of the effects presented in this paper deteriorate if the lattice filling is not perfect (not all of the lattice sites are occupied) and the trapping depth confining the atoms to the lattice is of finite magnitude (introducing uncertainty in the atomic positions). To model the finite trap depth, we assume the trapping potential is a standing wave of amplitude $V_0$ (considered to be approximately harmonic at the minima). The atomic wavefunctions are assumed to be those of ground state harmonic oscillators, centered on each lattice site. Each realization of the position is therefore determined according to a Gaussian probability distribution, ${\rho_i \propto \mathop{\exp (-[(x-x_i)^2 + (y-y_i)^2]/\ell^2) }}$, where $\ell=(a/\pi)(E_R/V_0)^{1/4}$ and $E_R$ is the lattice recoil energy (see Supplemental Materials in \cite{Jenkins2012,Bettles2015a} for further details). For relatively high filling factors ($90\%$ occupation) and significant trap depths $V_0 =750 E_R$, we see in Fig.\ \ref{Fig:4}(a) that the narrow subradiant modes responsible for the weak Fano resonances are washed out, leaving contributions from the broader, stronger modes only. Using the same lattice parameters in the kagome lattice however [Fig.\ \ref{Fig:4}(b)], the interference lineshape is still very clear to see. $90\%$ filling has recently been realized for a $2\times 2$ array \cite{Lester2015a} and trap depths of $10^3E_R$ are possible in, e.g., optical lattices \cite{Bakr2009} where high filling factors are possible via the Mott-insulator phase and algorithmic cooling \cite{Bakr2011}.

\section{Conclusions}
In conclusion, we have shown that dipoles arranged in a periodic 2D lattice with spacing of order of the driving wavelength respond cooperatively rather than independently. We observe cooperative decays and shifts akin to those predicted for pairs \cite{Czarnik1969,Wang2010} and 1D chains \cite{Nienhuis1987,Olmos2013} of atomic dipoles, with different superradiant and subradiant cooperative modes being populated. The interference of these modes produces non-trivial asymmetric line-shapes. A particularly striking example is shown in the kagome lattice where we observe cooperative EIT in a system with only 2 levels. This 2-level cooperative EIT corresponds to interlaced ferroelectric and antiferroelectric phases of the coupled spin system. 
These 2D lattices provide us with an exciting means to explore interesting many-body spin models as a test-bed for driven dissipative non-equilibrium systems, including in the quantum regime \cite{Zhu2015}, and may have direct applications in e.g.\ narrow linewidth optical lattice clocks \cite{Kramer} and subradiant quantum information storage \cite{Olmos2013}.

\begin{acknowledgements}
We acknowledge funding from the UK EPSRC (Grant No.\ EP/L023024/1), and thank I. G. Hughes, R. M. Potvliege, I. Lesanovsky, B. Olmos, M. D. Lee, J. Ruostekoski, S. Jenkins, M. Greiner and A. Browaeys for useful discussions. The data presented in this paper are available at DOI:\href{http://dx.doi.org/10.15128/kk91fm26x}{10.15128/kk91fm26x}.
\end{acknowledgements}


%


\end{document}